\documentclass[prc,showpacs,twocolumn]{revtex4}

\usepackage{graphicx}% Include figure files
\usepackage{epsfig}
\usepackage{dcolumn}% Align table columns on decimal point
\usepackage{bm}% bold math
\usepackage{amssymb}
\newcommand{\be}{\begin{equation}}
\newcommand{\ee}{\end{equation}}
\newcommand{\bea}{\begin{eqnarray}}
\newcommand{\eea}{\end{eqnarray}}

\begin{document}

\preprint{APS/123-QED}

\title{Zipf's law in Nuclear Multifragmentation and Percolation Theory}

\author{Kerstin Paech}
\author{Wolfgang Bauer}
\author{Scott Pratt}
\affiliation{%
Department of Physics and Astronomy\\
Michigan State University, East Lansing, MI 48824-2320, USA\\
}%

\date{\today}% It is always \today, today,

\begin{abstract}
We investigate the average sizes of the $n$ largest fragments in nuclear
multifragmentation events near the critical point of the nuclear matter phase diagram.
We perform analytic calculations employing Poisson statistics as well as Monte Carlo simulations
of the percolation type. We find that previous claims of manifestations of Zipf's Law
in the rank-ordered fragment size distributions are not born out in our result, neither in finite nor
infinite systems.  Instead, we find that Zipf-Mandelbrot distributions are needed to describe
the results, and we show how one can derive them in the infinite size limit.
However, we agree with previous authors that the investigation of rank-ordered
fragment size distributions is an alternative way to look for the critical point in the nuclear matter
diagram.
\end{abstract}

\maketitle

\section{Introduction}

A central goal of nuclear physics is to study the phase diagram of nuclear and quark-gluon
matter. Of special interest are two phase transitions, both of which are expected to be first-order transitions terminating at a critical point in the $\mu-T$ plane. At high temperatures ($T\approx 200$ MeV) and/or densities QCD tells us that
nucleons are no longer bound, but dissolve into assymptotically free
quarks and gluons \cite{Collins:1974ky}, possibly with an accompanying restoration of 
chiral symmetry \cite{Wilczek:2000ih}. The SPS-CERN and RHIC-BNL experiments 
have found evidence of this phase transition, and that is might be of first-order (or possibly a cross-over)\cite{Harris:1996zx,Muller:2006ee}. Theoretically, lattice-QCD calculations
suggest that the nuclear matter phase diagram contains a critical point at high temperature and density
\cite{Weber:2000xd,Fodor:2001pe,Fodor:2004nz}. One of the central goals of the future FAIR facility at the
German GSI is to study this critical point.

The second transition is under far better control, both experimentally and theoretically, and involves the transition of the low-temperature nuclear Fermi liquid into a gas of nucleons and small fragments. Experimentally 
\cite{Hirsch:1984yj,Gilkes:1994gc,Ritter:1994yt,Elliott:2000am,Lefort:1999si,Beaulieu:1999jv,Beaulieu:2001ra}, one finds 
a critical point at which this first-order transition terminates.
This critical point is located at temperature and 
density~\cite{KleineBerkenbusch:2001kq,BAP05,Bau07}
\begin{equation}
	T_c=8.3\pm0.2\ {\rm MeV},\ \ \rho_c/\rho_0=0.35 \pm 0.1
\end{equation}

In addition, through comparisons with phase transition theories, the set of
critical exponents determining the universality class of the phase transition
has been determined~\cite{KleineBerkenbusch:2001kq}. The experimental values of the exponents, 
extracted in
a somewhat model dependent way, are in good agreement with the universality class
of percolation~\cite{BDM85,Bauer:1986pv,Cam86,BKR86,NBD86,Bau88},
provided that appropriate finite size corrections are 
included~\cite{bauer-1995-52,Bauer:1997bw,KleineBerkenbusch:2001kq}. In passing we note that similar fragmentation patterns pointing to the existence of phase transitions
can also be found in the disintegration of molecules, such as buckyballs \cite{buck1,buck2}. With the
advent of the very high-powered free-electron lasers at DESY in Hamburg and at SLAC in Stanford
a fresh look at these molecular fragmentation experiments will be possible, and first experiments
have been proposed \cite{Berrah07}.

The great majority of analyses of critical phenomena in nuclear fragmentation have considered the shape of the mass distribution, which behaves as a power law in the critical region. In the present article we wish to address whether studying the behavior of the ranked fragment sizes, i.e., the sizes of the largest fragment, second largest, and so on, also contains information regarding critical behavior. As a second goal, we also investigate the applicability Zipf's law, an empirical expression sometimes linked to self-ordered criticality which has recently been applied to nuclear multifragmentation.

\section{Zipf's Law in Nuclear Multifragmention}

When Watanabe \cite{Wat96} and Ma et al.~\cite{Ma:1999qp,Ma:2004ey} studied the average size 
of the $n^{\rm th}$ largest fragment 
$\langle A_n \rangle$ as a function of $n$ they 
found that the data is described by Zipf's law~\cite{zipf} 
which was first found in
linguistics and is named after the linguist
George Zipf). Zipf's law is empirical, and
states that the most frequent word 
in a given language appears twice
as often as the second most frequent word, three times as often as the third most
frequent and so on, i.e. 
\begin{equation}
F(r)\propto r^{-1} \quad ,
\label{zipf}
\end{equation}
where $F(r)$ is the frequency of the $r$-th most frequent word.
A more general form,
\begin{equation}
F(r)\propto r^{-\lambda}
\label{zipf_lam}
\end{equation}
is also often refered to as Zipf's Law.

Since it was first formulated, many examples outside of linguistics
have shown the same behavior. It should also be noted that the Pareto
distribution~\cite{pareto:97}
(named after the economist Vilfredo Pareto) contains Zipf's law in a different
formulation and was also found in a broad range of statistical data. 

Ma et al. claim that since Zipf's law has been observed
in very different fields this is a reflection of self-ordered
criticality. Therefore, in~\cite{Ma:2004ey} Zipf's law in its 
extended form~(\ref{zipf_lam})
was fitted to
experimental data for the average size of the $n^{\rm th}$ largest fragment, see 
Figs.~23 and 24 in~\cite{Ma:2004ey}. 
They found that for increasing excitation energies
$\lambda$ decreases and passes through $\lambda=1$ at some point, i.e.\ 
Zipf's law in its original version~(\ref{zipf}). It is hoped
that this passing through $\lambda=1$ is an indication for the critical point.

%%%%%%%%%%%%%%%%%%%%%%%%%%%%%%%%%%%%%%%%%%%%%%%%%%%%%%%%%%%%%%%%%%%%

\section{$n$-th largest clusters and average sizes}
\label{sec_poisson}

%%%%%%%%%%%%%%%%%%%%%%%%%%%%%%%%%%%%%%%%%%%%%%%%%%%%%%%%%%%%%%%%%%%%

If we assume that the
production of any two clusters is largely independent, 
then the probability that a cluster of size $A$ is 
the $n^{\rm th}$-largest cluster
is given by
\begin{equation} 
P_{n}(A) = \sum_{i=0}^{n-1}
p_{\geq n-i}(A)
\cdot p_i(>A)~,
\label{p_nth}
\end{equation}
where 
\begin{equation}
p_{\geq m}(A) = 1 - \sum_{k=0}^{m-1} p_k(A) \quad 
\end{equation}
is the probability to have at least $m$ clusters of size $A$.

Here, $p_{k}(A)$ is the probability to have $k$ clusters of size
$A$ and $p_i(>A)$ is the probability that there are $i$ clusters of size
larger than $A$.
For example, the probability that a cluster is the largest in a given 
fragmentation event is the product of the probability $p_{\ge 1}(A)$
that there is at least 
one cluster of size $A$ present  and the probability $p_0(>A)$ 
that there are zero clusters of size 
larger than $A$
\begin{eqnarray}
P_{1}(A) &= &p_{\ge 1}(A)\cdot p_0(>A)\nonumber\\
&=& (1-p_0(A)) \cdot p_0(>A)
\end{eqnarray}
The probability that a cluster is the second largest is given by a sum of two terms. 
The first term is the probability $p_{\ge 2}(A)$ that there are two or 
more clusters of size $A$ present multiplied with
the probability  $p_0(>A)$ that there
are no clusters of size larger than $A$. The second term is the probability
$p_{\ge 1}(A)$ that there is at least one cluster of size $A$ present multiplied
with the probability $p_1(>A)$ that there is exactly one cluster of size
larger than $A$ present. This yields 
\begin{eqnarray}
P_{2}(A) & = & p_{\ge 2}(A) \cdot p_0(>A) +p_{\ge 1}(A)\cdot p_1(>A) \nonumber\\
&=& (1-p_0(A)-p_1(A)) \cdot p_0(>A) \nonumber\\
&& + (1-p_0(A)) \cdot p_1(>A)
\end{eqnarray}
for the probability that a cluster is the second largest in a given
fragmentation event.
From the probability $P_{n^{\rm th}}$
that a cluster of size $A$ is the
$n^{\rm th}$ largest 
we can calculate the average size of the $n$-th largest cluster
\begin{eqnarray}
\langle A_{n}\rangle = 
\sum_{A=1}^{V} A \cdot  P_{n} (A) \quad .
\label{mean_an}
\end{eqnarray}
Here, $P_{n} (A)$ depends on the probability distribution 
$p_i$ that is chosen 
for the underlying physical system. Since  we assume that the
production of any two clusters is largely independent from one another, 
the probability distribution can be approximated by the
Poisson distribution
\begin{equation}
p_k(A)=\frac{1}{k!} (N(A))^k \, e^{- N(A) } ~ ,
\label{poisson_prob}
\end{equation}
where $N(A)$ is the appropriate cluster size distribution,
i.e. the average of clusters of size $A$,  and may depend
on additional parameters like the excitation energy of the reaction.

For a system close to the critical point, the cluster size distribution 
follows a scaling function of the form
\begin{equation}
n(A,\epsilon) = a A^{-\tau} f(\epsilon A^\sigma) \quad ,
\end{equation}
where $\epsilon$ is the fractional deviation of the control parameter (for 
example the temperature $T$, i.e.\ $\epsilon=(T-T_{\rm c})/T_{\rm c}$) from
the critical value. The scaling function $f$
is equal to 1 at the critical point ($f(0)=1$). The critical exponents
$\tau$ and $\sigma$ determine the universality class of the phase transition, and the normalization constant $a$
is defined by the condition that all nucleons belong to some cluster
\begin{equation}
\sum_{A=1}^{V} A \cdot n(A,\epsilon) = V \quad ,
\label{def_a}
\end{equation}
where $V$ is the total number of nucleons in an event.
 
In~\cite{Bauer:2005bq} and~\cite{Campi:2005gc} 
the behavior of the average sizes of the $n$-th largest
cluster at the critical point ($\epsilon=0$) was studied assuming that 
the cluster sizes can
be described by a Poissonian probablity distribution. Here, we summarize those results and consider the critical  point 
where the cluster size distribution is a pure power law
\begin{equation}
N(A)=n(A,0)=aA^{-\tau}~,
\label{na_crit}
\end{equation}
which with equation (\ref{poisson_prob}) can be used to calculate
the average size of the $n$-th largest cluster according to equations (\ref{p_nth})
and (\ref{mean_an}) and we will consider $2 < \tau < 3 $ in the following. 
The normalization constant is given by
\bea
a = V / \sum_{A=1}^{V} A^{-(1-\tau)} 
= V/H_{V,1-\tau}~,
\eea
where 
\mbox{$H_{n,m}=\sum_{k=1}^{n} k^{-m}$}
is the $n^{\rm th}$ harmonic number of order m.
For the probability to have $i$-clusters larger than $A$,
\bea
N(>A) &=& \sum_{k=A+1}^{V}  N(k) 
= a   \sum_{k=A+1}^{V} k^{-\tau}\\
&=& a\left[ 
\zeta(\tau,1+A) - \zeta(\tau,1+V)
\right]
\nonumber
\label{nlarger}
\eea
has to be used with equation (\ref{poisson_prob}), where 
\mbox{$\zeta(s,q) = \sum_{k=0}^{\infty} (k+q)^{-s}$} 
is the generalized Riemann function.

For large systems, one can also replace the sum in 
equations~(\ref{def_a}) to~(\ref{nlarger})
by an integral. 
The constant $a$ is then defined by
\bea
 \int_{A_{\rm min}}^V {\rm d}  A \,\,
 a \,  A^{-(\tau-1)} & =& V
\eea
and is given by
\bea
a = V (\tau-2) \left(  A_{\rm min}^{-(\tau -2)} - V^{-(\tau -2)}\right)^{-1}~.
\nonumber
\eea

For the average number of fragments larger 
then $A$ we obtain
%\be
\bea
\eta(A) &=& \int_A^V {\rm d} \widetilde A \,\, a \, \widetilde A^{-\tau} \\
&=& \frac{a}{\tau -1} \left( A^{-(\tau -1)} - V^{-(\tau -1)}\right)
~ .
\nonumber\\
\label{eqn_eta_larger}
\eea
The probability that a cluster is the $n^{\rm th}$ largest is 
\be
P_{n}(A) = \frac{\eta^{n-1}}{(n-1)!}e^{-\eta} A^{-\tau}~,
\ee
and therefore the average size of the $n^{\rm th}$ largest cluster is
\be
\langle A_{n}\rangle =
\int_{A_{\rm min}}^V {\rm d}  A \,\,
 A \, a \,  A^{-\tau}
\frac{\eta^{n-1}}{(n-1)!}e^{-\eta} \, .
\ee
We can rewrite this integral by integrating over $\eta(A)$
with
\be
A(\eta)= \left(
\frac{\eta(\tau-1)}{a}
+ V^{-(\tau-1)}
\right)^{-\frac{1}{\tau-1}}~,
\ee
and get
\bea
\langle A_{n}\rangle &=&
\left(\frac{a}{\tau-1}\right)^\frac{1}{\tau-1}   
\int_{0}^{\eta_{\rm max}} {\rm d}\eta
\left(\eta + a \frac{V^{-(\tau-1)}}{\tau-1}\right)^{-\frac{1}{\tau-1}} 
\nonumber\\
& & \hspace{2cm} \times\quad
 e^{-\eta} \frac{\eta^{n-1}}{(n-1)!} \quad ,
\eea
where $\eta(V)=0$ and $\eta_{\rm max}  = \eta(A_{\rm min})$.
For average size of the $n+1$-th largest fragment in an infinite system 
this leads to
\bea
\langle A^\infty_{n}\rangle =
\left(\frac{a}{\tau-1}\right)^\frac{1}{\tau-1}   
\int_{0}^{\infty} {\rm d}\eta
\, e^{-\eta} \,\frac{\eta^{n-\frac{1}{\tau-1}-1}}{(n-1)!}
\eea
which we can express in terms of the gamma function as
\bea
\langle A^\infty_{n}\rangle&= &
\left(\frac{a}{\tau-1}\right)^\frac{1}{\tau-1}
\frac{\Gamma(n-\frac{1}{\tau-1})}{(n-1)!} 
\nonumber
\\
&=&
\left(\frac{a}{\tau-1}\right)^{\frac{1}{\tau-1}}
\frac{\Gamma\left(n-\frac{1}{\tau-1} \right)}{\Gamma(n)}
\quad ,
\label{mean_an_cont}
\eea
where in the last step we made use of the fact that $\Gamma(n)=(n-1)!$.

\begin{figure}[h]
\centerline{\includegraphics[width=9cm]{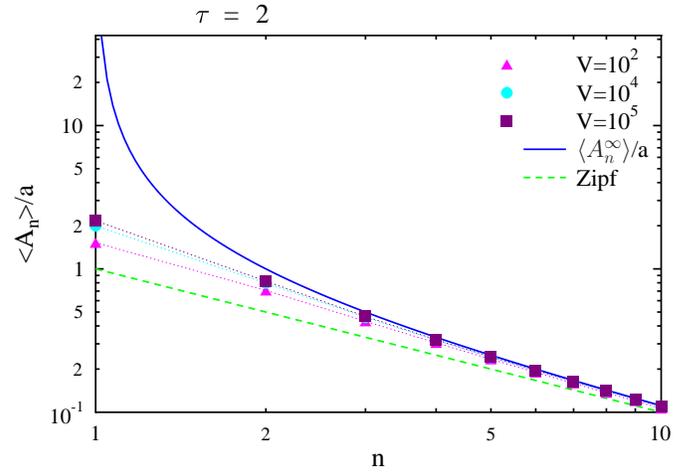}}
%\epsfig{figure=an_poisson.eps ,width=9cm}
%\vspace*{-2cm}
\caption[]{%
Symbols indicate $\langle A_{n}\rangle$ for $\tau=2$ and
different system sizes 
calculated with equation~(\ref{mean_an}), the solid line indicates 
$\langle A^\infty_{n}\rangle$ from equation~(\ref{an_tau2}), the
dashed line indicates Zipf's law~(\ref{zipf}).
}
\label{fig_an_poisson}
\end{figure}

Hence, in the case of an infinite system for $\tau=2$ the average size 
of the $n^{\rm th}$ largest cluster
is 
\be
\langle A^\infty_{n}\rangle = a \, 
\frac{\Gamma\left( n-1\right)}{\Gamma(n)} = a \, (n-1)^{-1}
\quad,
\label{an_tau2}
\ee
where we used the identity 
$\Gamma(x+1)=x\Gamma(x)$. Therefore for $\tau=2$ the average size of 
the  $n^{\rm th}$ largest cluster in an infinite system 
is not described by Zipf's law, but
by the Zipf-Mandelbrot distribution~\cite{citeulike:580392,Mandelbrot:CT:1953}
\begin{equation}
F(r)=c(r+k)^{-\lambda}
\label{zipf_mandel}
\end{equation}
with $k=-1$ and $\lambda=1$.
Fig.~\ref{fig_an_poisson} shows $\langle A_{n}\rangle$ calculated with equation~(\ref{mean_an}),
divided by the normalization constant $a$
for different system sizes $V$. For $n=1$ and $n=2$ the finite size effects are very noticable,
however for $n \ge 3$ the $\langle A_{n}\rangle$ match the Zipf-Mandelbrot distribution. 
It should
also be noted that Zipf's law with $\lambda=1$ does not describe $\langle A_{n}\rangle$
very well.

Using the ansatz described above we determine  $\langle A_{n}\rangle$ for different
values $2\le \tau \le3$ and test how well they are described
by Zipf's law and the Zipf-Mandelbrot distribution. 

\begin{figure}[h]
\centerline{\includegraphics[width=9cm]{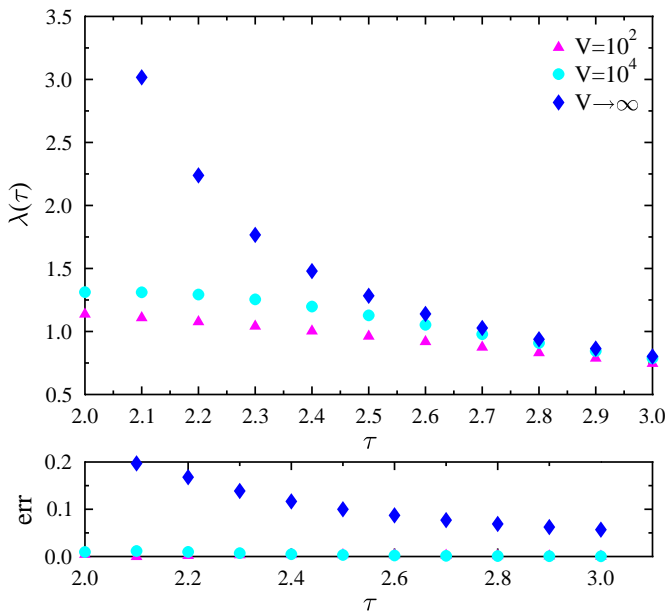}}
%\epsfig{figure=lam_z.eps,width=9cm}
%\vspace*{-2cm}
\caption[]{%
Top: $\lambda$ as a function of $\tau$ as 
fitted to Zipf's law~(\ref{zipf_lam})
for different system sizes.
Bottom: Error of the fit
}
\label{fig_lam_z}
\end{figure}

Fig.~\ref{fig_lam_z} shows the extracted fits of the parameter 
$\lambda$ for Zipf's law for different values $\tau$ and different 
system sizes $V$.
The value of $\lambda$ for the finite systems shows a 
decrease from $\lambda(\tau=2)=1.25$ to $\lambda(\tau=3)\approx 0.75$,
whereas for the limit of an infinitely large system there is a
stronger dependence on $\tau$. The errors of the fit are 
shown in the lower part of the figure, demonstrating that the fits for
the finite systems are a good approximation, while the error for the
infinite system is substantially higher.
For the finite systems, $\lambda$ is relatively close to unity
for the small system with $V=10^2$ around $\tau=2.5$. However, for example
$\lambda(\tau=3)\approx 0.75$ which would indicate that the statement
of~\cite{Ma:2004ey} that $\lambda=1$ as found in the experimental data
is more coincidental than profound. Fig.~\ref{fig_lam_z} shows
that the value of $\lambda$ at the critical point depends on $\tau$ 
and the system size $V$ and that $\lambda=1$ at the critical point
only holds true for certain sets of parameters.

\begin{figure}[h]
\centerline{\includegraphics[width=9cm]{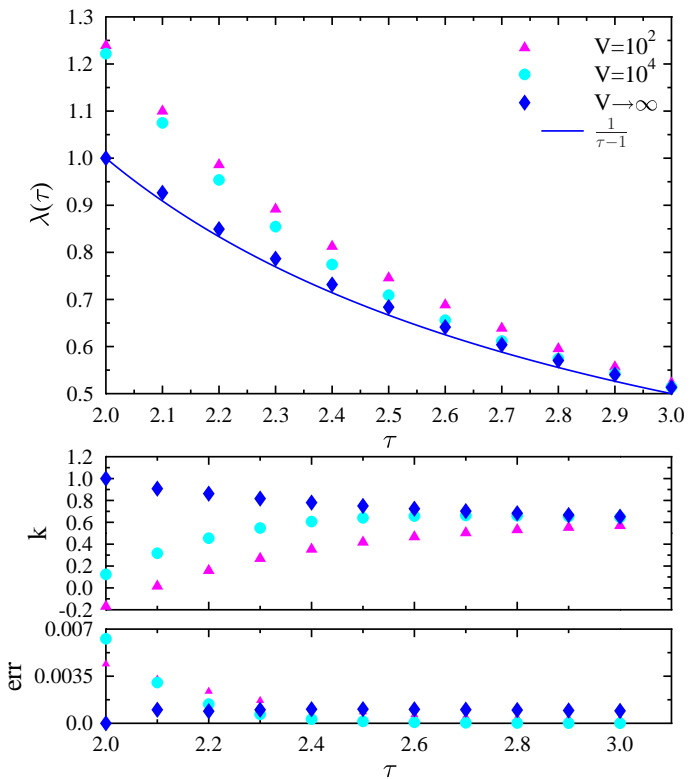}}
%\epsfig{figure=lam_zm.eps,width=9cm}
%\vspace*{-2cm}
\caption[]{%
$\lambda$ as a function of $\tau$ as 
 fitted to the Zipf-Mandelbrot distribution~(\ref{zipf_mandel})
for different system sizes. 
Middle: Extracted parameter $k$.
Bottom: Error of the fit
}
\label{fig_lam_zm}
\end{figure}
Fig.~\ref{fig_lam_zm} shows the extracted fits of the parameter 
$\lambda$ for the Zipf-Mandelbrot distribution
 for different $\tau$ and different 
system sizes $V$. The value of $\lambda$ for the finite systems shows a 
decrease from $\lambda(\tau=2)\approx 1.2$ to $\lambda(\tau=3)\approx 0.5$,
while for the larger system the decrease is not as strong. The 
difference between the finite systems and the infinite system is not
as strong as for the fit to Zipf's law. 
The errors of the fit are 
shown in the lower part of the figure showing that the fits for
the finite systems are indeed 
a much better approximation of the $\langle A_{n}\rangle$
than Zipf's law,
and the fit for the infinite system yields the best approximation
overall.
The parameter $k$ shown in Fig.~\ref{fig_lam_zm} (middle) shows
some finite size dependency for small $\tau$, however the 
system size dependence decreases with increasing $\tau$.

From fits to the Zipf-Mandelbrot distribution  
for $2\le \tau \le 3$
one finds a systematic 
dependence of $\lambda$ on the critical exponent $\tau$ of the form
\be
\lambda \approx \frac{1}{\tau-1} \quad ,
\ee
which holds true almost exaclty for the infinite system and
is a good approximation for small systems.

To summarize the findings of this section: Zipf's law should not be
applied to nulcear fragmentation, but rather the more general 
form of the Zipf-Mandelbrot distribution. And $\lambda=1$ as 
found in the experimental data
is more coincidental than a measure of a critical exponent or an indicator of the self-similar behavior.

%%%%%%%%%%%%%%%%%%%%%%%%%%%%%%%%%%%%%%%%%%%%%%%%%%%%%%%%%%%%%%%%%%%%

\section{Percolation theory and Zipf's law}

%%%%%%%%%%%%%%%%%%%%%%%%%%%%%%%%%%%%%%%%%%%%%%%%%%%%%%%%%%%%%%%%%%%%

The ansatz described in~\ref{sec_poisson} can only be applied 
for systems at (or close to) the critical point since information 
about the scaling function $f$ is
needed for $\epsilon \ne 0$. It has been discussed in~\cite{Bauer:1986pv} that
bond percolation can be used to describe multi-fragmentation reactions. 
In the following we will use three dimensional bond percolation
model on a simple cubic lattice~\cite{grimmett89percolation,Stauffer92}.
The probability to form a bond between two neighboring lattice 
sites is $p_{\rm bond}$.

The critical exponents of the percolation model depend on the system size 
and the dimensionality of the lattice. For an 
infinitely large three dimensional lattice $\tau=2.18$ and $\sigma=0.45$. 
The critical bond probability does not only depend on the dimensionality of a
lattice, but also on its particular topology. For an infinitely large simple cubic 
lattice the critical probability is $p_{\rm c}=0.249$.

In the following, we will consider two different system sizes:
a small lattice with 
$6\times 6\times 6 = 216$
sites (corresponding to system sizes found in nuclear fragmentation reactions)
and a large lattice with $100\times 100\times 100 = 10^6$ sites 
(reducing the finite size effects considerably).

\begin{figure}[h]
\centerline{\includegraphics[width=9cm]{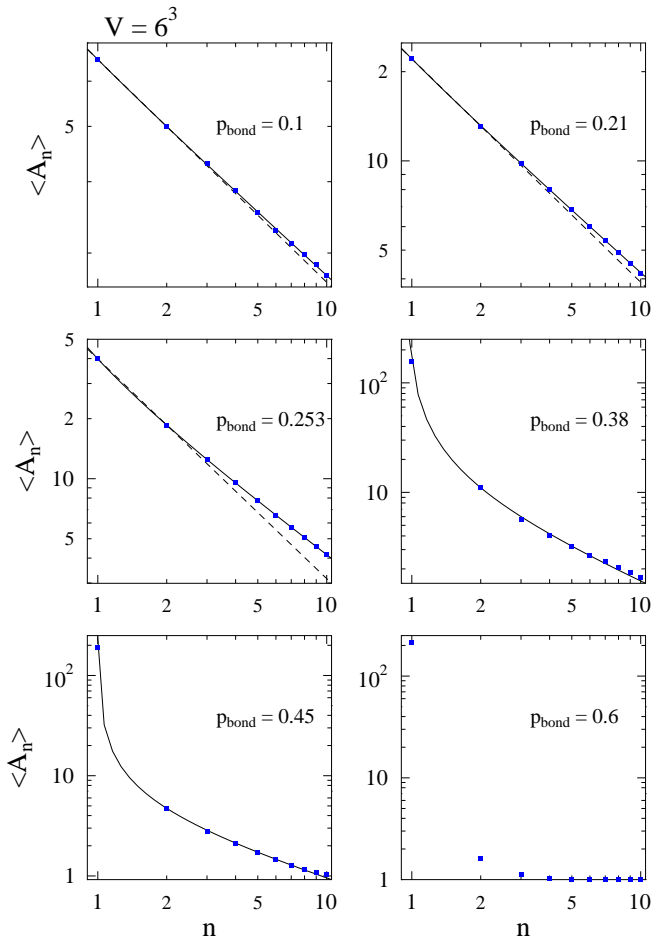}}
%\epsfig{figure=L6_an.eps,width=9cm}
%\vspace*{-2cm}
\caption[]{%
Average sizes of the $n$-th largest cluster
as a function of $n$ for the lattice with 216 sites for different
bond probabilities $p_{\rm bond}$. The dashed lines are fits for
Zipf's law~(\ref{zipf_lam}), the solid lines are fits for the 
Zipf-Mandelbrot distribution~(\ref{zipf_mandel}).
}
\label{an_pbond_6}
\end{figure}

Fig.~\ref{an_pbond_6} shows the average sizes of the $n$-th largest cluster
as a function of $n$ for the lattice with 216 sites for different
bond probabilities $p_{\rm bond}$. For this system size we effectively have
$\tau \approx 2.14$ at the critical point $p_{\rm c}\approx 0.253$. 
The value for $p_{\rm c}$ was chosen by determining the value of $p_{\rm bond}$ 
where the cluster size distributions
$N(A)$ were fitted best by the power law distribution~(\ref{na_crit}).

The results for $\langle A_n\rangle$ from percolation theory are similar to 
the experimental data shown in Fig. 23 from~\cite{Ma:2004ey}. 
The small $p_{\rm bond}$ correspond
to high excitation energies and vice versa. For our results from
percolation we 
see that the fits for Zipf's law (and Zipf-Mandelbrot) work very well below
$p_{\rm c}$ (that corresponds to excitation energies above the critical point).
But as one crosses the critical point to $p_{\rm bond}>p_{\rm c}$ 
(corresponding to excitation energies above the critical point) 
we observe that Zipf's law (and eventually Zipf-Mandelbrot) does not work if 
$\langle A_1\rangle$ is included.

We fit Zipf's law~(\ref{zipf_lam}) to $\langle A_n\rangle$ 
from percolation and determined
the parameter $\lambda$.  
Since we use a least squares fitting procedure we use two different 
methods to obtain a fit. The first method (denoted ``Fit 1'') 
is equivalent to what is done
for Fig. 24 in~\cite{Ma:2004ey}, i.e. fitting the data to equation~(\ref{zipf_lam})
with a least square fit. However, this means that for a power law fit
the first two data points will dominate the result for the fit. As a second
method (denoted ``Fit 2'') we first take the logarithm of both axis and
then fit the $\langle A_n\rangle$ to a straight line to determine $\lambda$,
hence all the $\langle A_n\rangle$ contribute more or less equally
to the fit result.

\begin{figure}[h]
\hspace*{-0.4cm}
\centerline{\includegraphics[width=9cm]{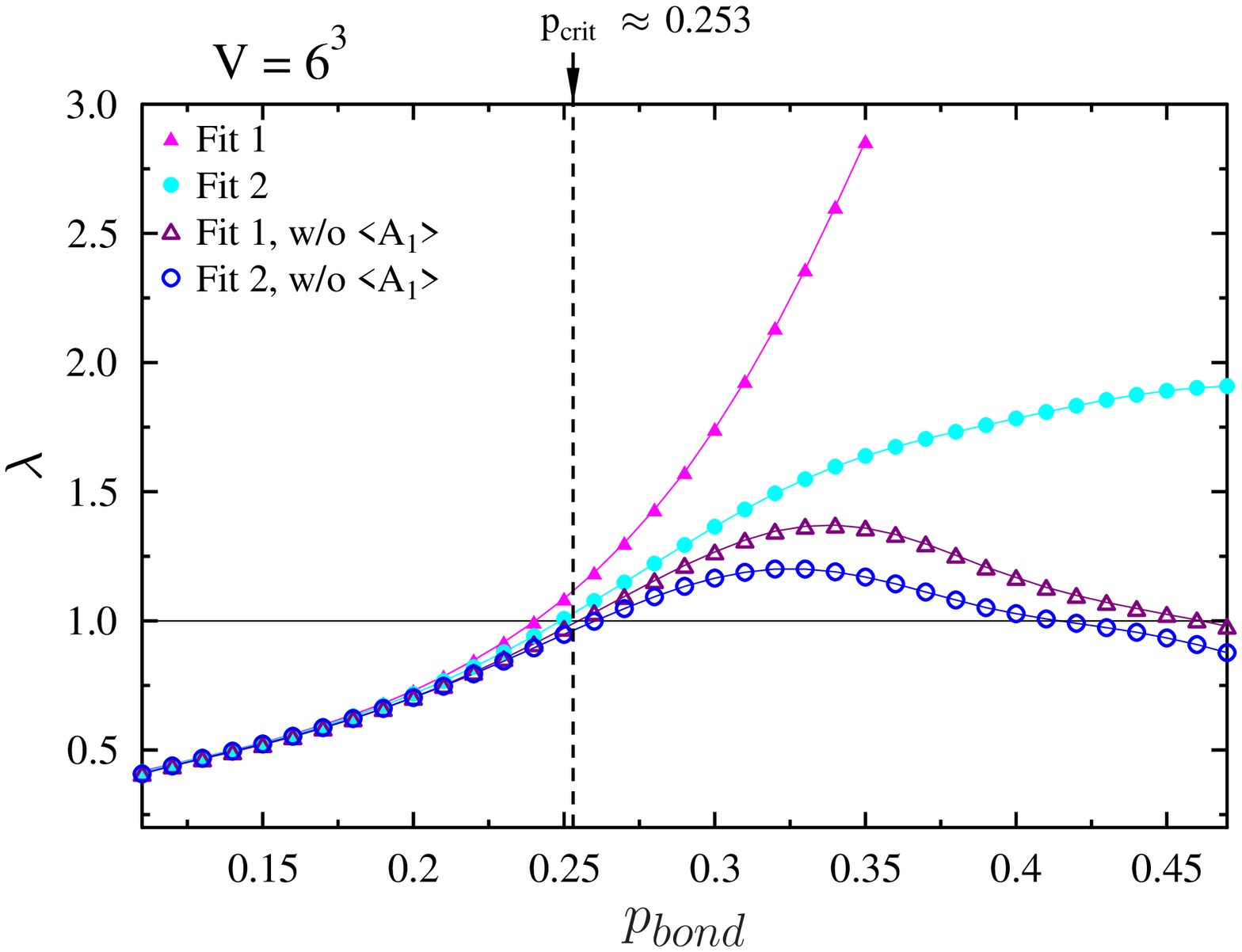}}
\hspace*{-0.4cm}
\centerline{\includegraphics[width=9cm]{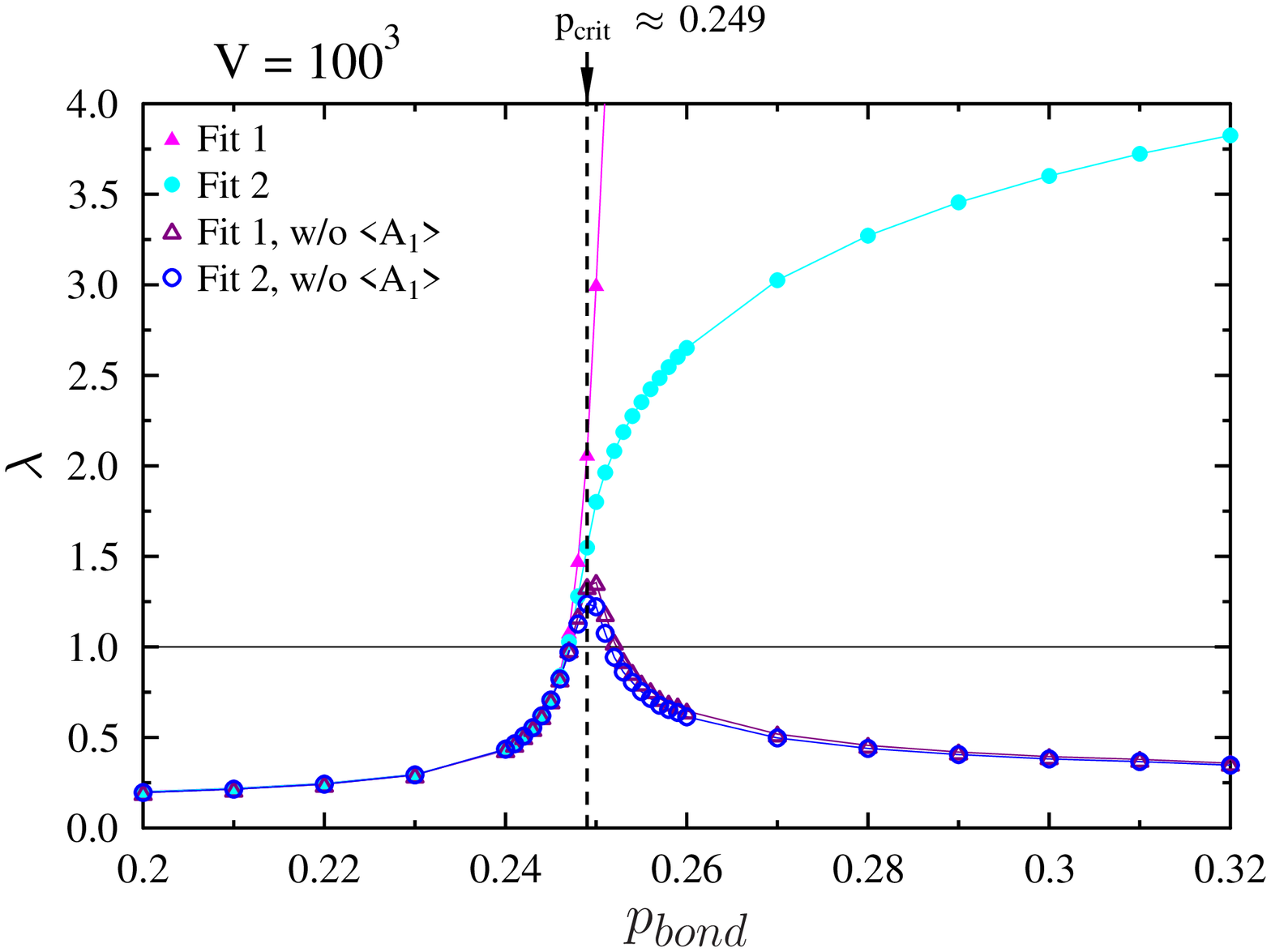}}
\caption[]{%
Fitted values for $\lambda$ using the different fits described in the text.
Solid symbols are for fits including the first fragment, open symbols
are for fits omitting $\langle A_n\rangle$. Upper figure is for
the small lattice (216 sites), lower figure is for the large lattice 
($10^6$ sites). Approximate values for $p_{\rm c}$ are also indicated by a
dashed line for
both cases.
}
\label{lam_6_100}
\end{figure}

We already mentioned that Zipf's law (Zipf-Mandelbrot) does not work 
if the average size of the largest fragement $\langle A_1 \rangle$
is included above $p_{\rm c}$. Therefore we perform the two fit
methods once including all the fragments and once excluding $\langle A_1 \rangle$
from the fit.
This way we have four different ways to determine a value for
$\lambda$ and the results are shown in Fig.~\ref{lam_6_100}, for both the
small and the large lattice.

For the small system we see a similar behavior as for the experimental data
from~\cite{Ma:2004ey} and  $\lambda(p_{\rm bond})$ crosses 
$\lambda=1$ at or very close to $p_{\rm c}$ for both system sizes
investigated. The fits are all very similar below $p_{\rm c}$ 
as the first fragment hardly deviates from Zipf's law.

In case of the large system ($10^6$ sites), and for the fit including
$\langle A_1\rangle$,  a quick rise of $\lambda$ can be observed with
increasing $p_{\rm bond}$ due to 
the domination of $\langle A_1 \rangle$. 
For the fit excluding $\langle A_1 \rangle$ we see a distinct
peak of $\lambda(p_{\rm bond})$ close to the critical point.

Although some maximum is 
visible for the case of the small system (216 sites), the maximum is
very broad and not particular near the critical point and can therefore 
not be
taken as an indication of the critical point for these small systems.

If we fit $\langle A_n\rangle$ with the Zipf-Mandelbrot 
distribution~(\ref{zipf_mandel}),
we get a similar behavior for $\lambda$. 
Further, for the case where we exclude  the largest fragment 
from the fit, the parameter $k$ also indicates the critical point
in the case of the large lattice ($10^6$ sites). Here we observe 
that $k\approx 0$ for $p_{\rm bond}<p_{\rm c}$ and a sharp drop to 
$k\approx -1$ for $p_{\rm bond}>p_{\rm c}$, i.e. the dependence
of $k$ can be approximated by
\begin{equation}
k(p_{\rm bond})\approx -\Theta(p_{\rm bond}-p_{\rm c}) \quad .
\end{equation} 
This means that we recover Zipf's law if we neglect the largest fragment
above $p_{\rm c}$.
However, when turning to the small lattice
($216$ sites), the sharp drop of $k$ at $p_{\rm c}$ turns into 
an extremely smooth decrease over a range in the bond probability 
of $\Delta p_{\rm bond}\approx 0.2$.

Therefore, the crossing through
$\lambda=1$ in the experimental data could indeed indicate
the crossing through the
critical point. 
However, we should keep in mind the results from sect.~\ref{sec_poisson}
which clarified that the value of $\lambda$ at the critical point
depends on $\tau$ and $V$.
Therefore, the fact that $\lambda=1$
close to the critical point cannot be taken as a unambigous signature
for the critical point.

\section{Percolation theory and $\langle A_n\rangle$}
We have argued in the previous section that the Zipf's law parameter $\lambda$ 
is not a rigorous signature for the critical point. In this section we want
to investigate what can be learned from considering the average
size of the $n$-th largest clusters $\langle A_n\rangle$.

We turn to look at the ratio of the $n$-th largest average cluster sizes 
$\langle A_{n+1}\rangle/\langle A_{n}\rangle$. If there is a critical 
behavior that indicates the critical point we hope to find it by looking
at the ratio of the $\langle A_n\rangle$ for different $n$.

\begin{figure}[h]
\hspace*{-0.4cm}
\centerline{\includegraphics[width=9cm]{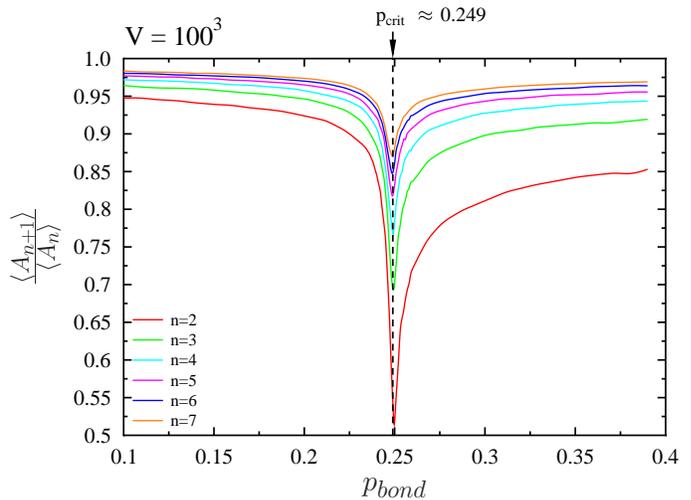}}
%\epsfig{figure=L100_anp1_an.eps,width=9cm}
%\vspace*{-2cm}
\caption[]{%
Ratio of the average size of the $n+1$-th largest fragment to 
average size of the $n$-th largest fragment as
a function of $p_{\rm bond}$ for a large system
with $10^6$ sites. The result for $n=1$ is omitted.
}
\label{pic_anp1_an_100}
\end{figure}

Fig.~\ref{pic_anp1_an_100} shows the ratio 
$\langle A_{n+1}\rangle/\langle A_n\rangle$ (with $n=2\dots 7$)
as a function of $p_{\rm bond}$ for the large system ($10^6$ sites).
Close to the critical point all ratios show a significant minimum at 
the critical bond probability $p_{\rm c}$. 
The average size of the largest fragment 
$\langle A_1\rangle$ 
has a very different behavior compared to $\langle A_n\rangle$ for 
$n>1$ and is therefore omitted.

Now we want to compare the results shown in Fig.~\ref{pic_anp1_an_100}
to the ansatz from section~\ref{sec_poisson} for an 
infinite system. In section~\ref{sec_poisson}
two assumptions were made: at the critical point the cluster size
distributions are discribed by a power law and the production of
the clusters is uncorrelated and therefore Poissonian.

For percolation theory we also expect to have a power law for the
cluster size distributions  at the critical point. 
Therefore, the only difference we
would expect at the critical point compared to the 
ansatz from section~\ref{sec_poisson} for an infinite system besides
finite size effects, would be due to correlations. 
These correletions would stem
from the fact that the production of large clusters is not 
independent. For example, if a single very large cluster is produced 
whose size is of the order of the total sytem size no further large 
clusters can be produced in the same event. However, the production
of the smaller clusters would not be affected.
Hence, at the critical point we wouldn't expect the Poissonian ansatz to 
work for the ratio $\langle A_{n+1}\rangle/\langle A_n\rangle$ for $n=1$, 
but possibly for $n$ larger than 2 or 3.

\begin{figure}[h]
\hspace*{-0.4cm}
\centerline{\includegraphics[width=9cm]{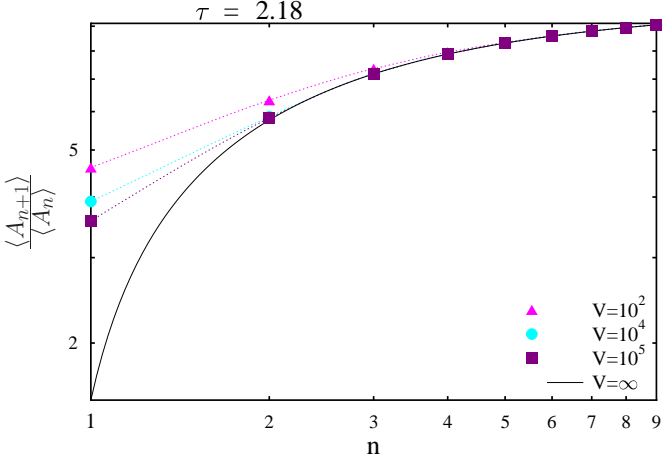}}
%\epsfig{figure=anp1an_poiss_tau218.eps,width=9cm}
%\vspace*{-2cm}
\caption[]{%
The ratio ${\langle A_{n+1}\rangle}/{\langle A_{n}\rangle}$
for different finite system sizes $V$ (symbols) using 
eqn.~(\ref{mean_an}) and for the infinte
system  (line) using eqn.~(\ref{mean_an_cont}) for $\tau=2.18$.}
\label{pic_anp1an_tau218_poisson}
\end{figure}

\begin{figure}[h]
\hspace*{-0.4cm}
\centerline{\includegraphics[width=9cm]{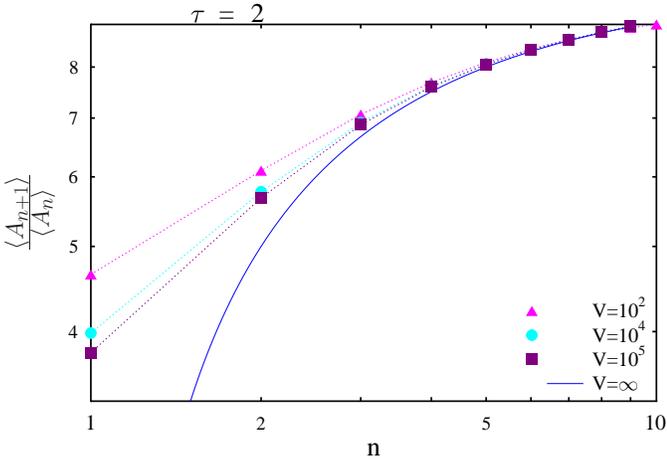}}
%\epsfig{figure=anp1an_poiss_tau2.eps,width=9cm}
%\vspace*{-2cm}
\caption[]{%
The ratio ${\langle A_{n+1}\rangle}/{\langle A_{n}\rangle}$
for different finite system sizes $V$ (symbols) using 
eqn.~(\ref{mean_an}) and for the infinte
system (line) using eqn.~(\ref{an_tau2}) for $\tau=2$.}
\label{pic_anp1an_tau2_poisson}
\end{figure}

Before comparing the percolation results to the 
ansatz from section~\ref{sec_poisson} for an infinite system
we first want to make sure that finite size effects are
negligible. Fig.~\ref{pic_anp1an_tau218_poisson} shows the
ratio ${\langle A_{n+1}\rangle}/{\langle A_{n}\rangle}$ for 
different system sizes.
It shows that finite size effects  between
the infinite system and the finite system sizes are 
negligible for $n>1$ for the system sizes larger
than $V=10^2$, and for $n>2$ are negligible for all shown
system sizes. Therefore, the 
ansatz from section~\ref{sec_poisson} for an infinite system
can be used for the comparison to the percolation model 
if the ratio for $n=1$ (and $n=2$ for the smaller system)
is excluded.
However, it should be noted that for $\tau=2$ the finite
size effects are again substantial for $n=1$ and 
non-negligible for $n=2$ and $n=3$, the finite size effects get
negligible for $n>3$,
as is shown in Fig.~\ref{pic_anp1an_tau2_poisson}

To compare the percolation result to the results from 
ansatz from section~\ref{sec_poisson}
that resulted in equation (\ref{mean_an_cont}).
Using the identity $\Gamma(x+1)=x\Gamma(x)$ we can write 
the ratio as
\begin{equation}
\frac{\langle A^\infty_{n+1}\rangle}{\langle A^\infty_{n}\rangle}
=\frac{n(\tau-1)-1}{n(\tau-1)}
\quad .
\label{anp1_an_poisson}
\end{equation}

\begin{figure}[h]
\hspace*{-0.4cm}
\centerline{\includegraphics[width=9cm]{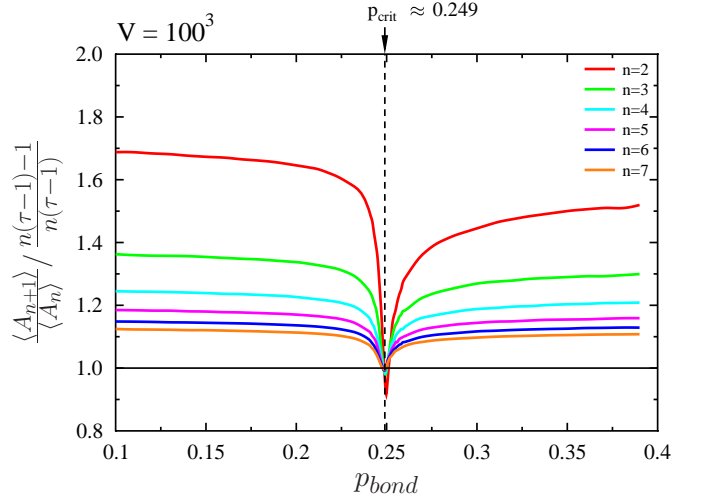}}
%\epsfig{figure=L100_norm.eps,width=9cm}
%\vspace*{-2cm}
\caption[]{%
Ratio of the average size of the $n+1$-th largest fragment to 
average size of the $n$-th largest fragment as
a function of $p_{\rm bond}$ normalized by the factor 
in equation~(\ref{anp1_an_poisson}) for a large system
($10^6$ sites). The result for $n=1$ is omitted.
}
\label{pic_anp1_an_100_norm}
\end{figure}
The ratio of the percolation result and the 
ansatz from section~\ref{sec_poisson} 
as a function of $p_{\rm bond}$ is
shown in Fig.~\ref{pic_anp1_an_100_norm}. 
The minima all come down to 1 for $n\ge 3$ and one can conclude 
that for the percolation model (at the critical point
where the fragments are distributed according to a power law) 
the production of the $n$-th largest cluster with $n\ge 2$ 
is not correlated for a large system. 

\begin{figure}[h]
\hspace*{-0.4cm}
\centerline{\includegraphics[width=9cm]{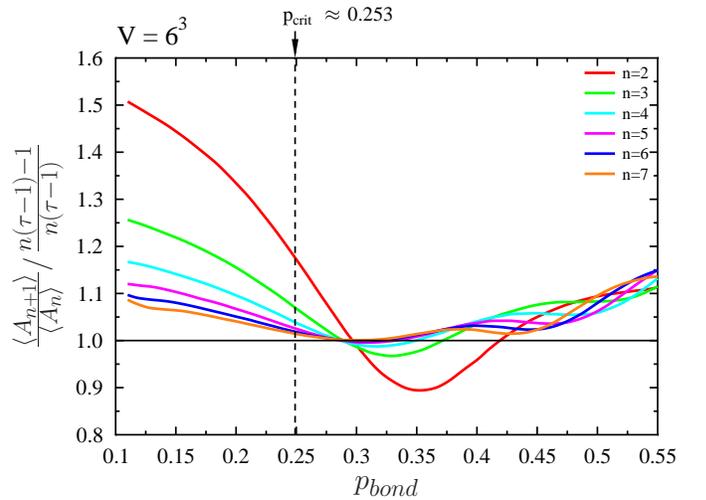}}
%\epsfig{figure=L6_norm.eps,width=9cm}
%\vspace*{-2cm}
\caption[]{%
Ratio of the average size of the $n+1$-th largest fragment to 
average size of the $n$-th largest fragment as
a function of $p_{\rm bond}$ normalized by the factor 
in equation~(\ref{anp1_an_poisson}) for a small system
($216$ sites). The result for $n=1$ is omitted.
}
\label{pic_anp1_an_6_norm}
\end{figure}

When turning to the small system (216 sites) shown in 
Fig.~\ref{pic_anp1_an_6_norm} however, the strong indication of
the critical point by the location of the pronounced minima does
not hold true anymore. We observe a broadening of the minima and
a shift of the minima away from the critical point towards higher
$p_{\rm bond}$. For $n\ge 5$ the minima still meet at some
$p_{\rm bond}$, but it is not located close to 
$p_{\rm c}\approx 0.253$.
This means that the percolation model and 
ansatz from section~\ref{sec_poisson}
agree for values of $p_{\rm bond}$ where the cluster distribution is no longer
a pure power law.

\section{Summary}

In contrast to previous claims, we have shown that a strict application of Zipf's
law to rank-ordered fragment size distributions near the critical point of the
nuclear fragmentation phase transition is not warranted.  This was done in analytical
calculations assuming independent Poisson emission probabilities as well as in
calculations based on percolation models, which include all correlations.  In order to
obtain information on finite-size effects, we have performed these calculations over
a wide variety of system sizes, and compared to analytical results for infinite systems. We have found that the rank-ordered fragment size distributions follow the more
general Zipf-Mandelbrot distributions instead of the simple Zipf law, and we have been able
to derive this relation in the infinite size limit for $\tau$ approaching 2.

For very large system sizes, we also find that monitoring the value of the Zipf-Mandelbrot
distribution fit to the rank-ordered fragment sizes yields a valuable signal for the detection of the
critical point in the phase diagram.  However, for the extremely small systems
typical of nuclear fragmentation we find that this signal is not nearly as reliable as the
previously explored scaling analysis \cite{Elliott:2001hn,KleineBerkenbusch:2001kq,Mader:2003re}.

\begin{acknowledgments}
This research was supported
by NFS grant PHY-0555893 (WB) and the
U.S. Department of Energy, Grant No. DE-FG02-03ER41259 (SP).
KP greatfully acknowledges support by the Feodor Lynen Program of 
the Alexander von Humboldt Foundation.
\end{acknowledgments}

%---> \bibitem{BAP05} W. Bauer, B. Alleman and S. Pratt, Proceedings for IWM 2005, Catania, Italy,
%	 ISBN 88-7438-029-1, (2005) 279.

\newpage %Just because of unusual number of tables stacked at end
\bibliographystyle{prsty}
\bibliography{zipf}

\end{document}